\documentclass[10pt,twocolumn,letterpaper]{article}

\usepackage[pagenumbers]{cvpr} % To force page numbers, e.g. for an arXiv version

\usepackage{graphicx}
\usepackage{amsmath}
\usepackage{amssymb}
\usepackage{booktabs}
\usepackage{graphicx}
\usepackage{float}
\usepackage{bm}
\usepackage{times}
\usepackage{caption}
\usepackage{makecell}
\usepackage{multicol}
\usepackage{multirow}
\usepackage{color}
\usepackage{booktabs}
\usepackage{float}
\usepackage{amssymb}
\usepackage{wasysym}
\usepackage[pagebackref,breaklinks,colorlinks]{hyperref}

\usepackage[capitalize]{cleveref}
\crefname{section}{Sec.}{Secs.}
\Crefname{section}{Section}{Sections}
\Crefname{table}{Table}{Tables}
\crefname{table}{Tab.}{Tabs.}

\def\cvprPaperID{382} % *** Enter the CVPR Paper ID here
\def\confName{CVPR}
\def\confYear{2022}

\begin{document}

\title{Transformations in Learned Image Compression  
\\from Modulation Perspective}

\author{Youneng Bao$\displaystyle ^{1}$, Fangyang Meng$\displaystyle ^{2}$, Wen Tan$\displaystyle ^{1}$, Chao Li$\displaystyle ^{1}$, Yonghong Tian$\displaystyle ^{2}$, Yongsheng Liang$\displaystyle ^{1,*}$ \\
$\displaystyle ^{1}$Harbin Institute of Technology, Shenzhen\ \ \ $\displaystyle ^{2}$Peng Cheng Laboratory \\
{\tt\small $\displaystyle ^{1}$\{baoyouneng, tanwen150548, lcc2332021\}@163.com  \ $\displaystyle ^{1}$liangys@hit.edu.cn \  $\displaystyle  ^{2}$\{mengfy, tianyh\}@pcl.ac.cn }
}
% \and
% Fanyang Meng\\
% \and
% Wen Tan\\
% {\tt\small tanwen150548@163.com}
% \and
% Chao Li\\
% \and
% Yousheng Liang^{*}\\
\maketitle

%%%%%%%%% ABSTRACT
\begin{abstract}
 
In this paper, a unified transformation method in learned image compression(LIC) is proposed from the perspective of communication. Firstly, the quantization in LIC is considered as a generalized channel with additive uniform noise.
Moreover, the LIC is interpreted as a particular communication system according to the consistency in structures and optimization objectives. Thus, the technology of communication systems can be applied to guide the design of modules in LIC. Furthermore, a unified transform method based on signal modulation (TSM) is defined. In the view of TSM, the existing transformation methods are mathematically reduced to a linear modulation. A series of transformation methods, e.g. TPM and TJM, are obtained by extending to nonlinear modulation. The experimental results on various datasets and backbone architectures verify that the effectiveness and robustness of the proposed method. More importantly, it further confirms the feasibility of guiding LIC design from a communication perspective. For example, when backbone architecture is hyperprior combining context model, our method achieves 3.52$\%$ BD-rate reduction over GDN on Kodak dataset without increasing complexity. 

\end{abstract}

%%%%%%%%% BODY TEXT
\section{Introduction}
\label{sec:intro}

Image and video compression is an important and fundamental problem for many years to enable efficient storage and transmission. Classical image compression standards rely on carefully hand-designed components to present module-based codec, such as JPEG\cite{marcellin2000overview}, JPEG2000\cite{rabbani2002overview}, HEVC/H.265\cite{sullivan2012overview} and VVC/H.266\cite{bross2021overview}. Since the modules are optimized independently, it is difficult to obtain optimized results for different objectives\cite{Cui_2021_CVPR}. With the rapid development of deep learning techniques, there has been a lot of works exploring end-to-end optimized image compression frameworks based on artificial neural networks\cite{balle2017end, balle2016endpcs, cai2018deep}. In recent years, the performance of learned based image compression frameworks\cite{hu2020coarse,cheng2019learning,balle2020nonlinear,li2018learning} have outperformed classical image compression codecs.\par

\begin{figure}[t]
\begin{center}

\includegraphics[width=8.5cm]{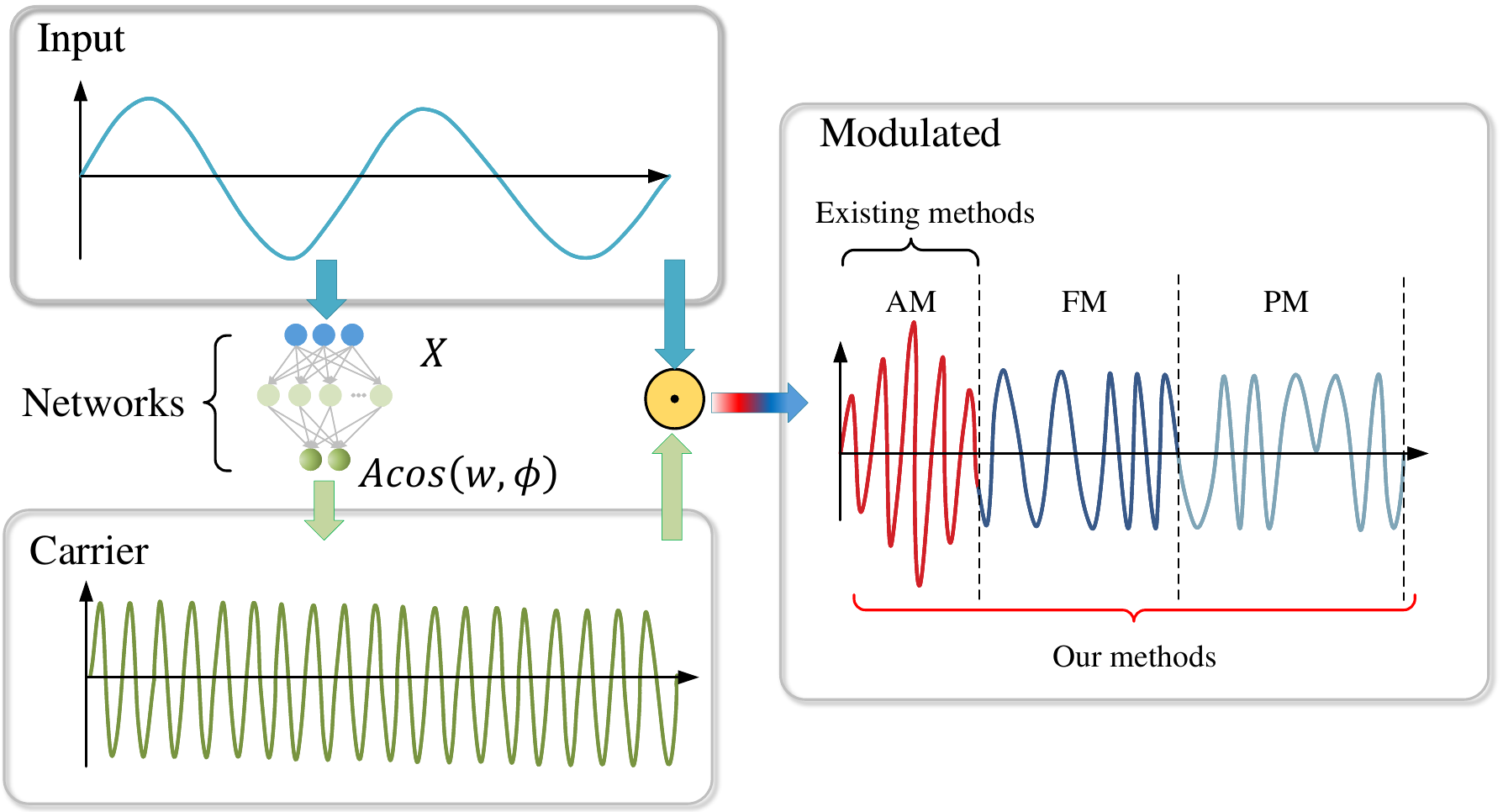}
\end{center}
\vspace{-1.5em}
\caption{The transformations in learned image compression are considered as the modulation of the signal. Existing methods can be classified as linear modulation. By extending to nonlinear modulation, a series of transformation methods are yielded.}
\vspace{-1em}
\end{figure}

The common keys of these currently most successful approaches are entropy modeling and transformation network based on the variational autoencoder (VAE) structure\cite{kingma2013auto}. Transformation can reduce the correlation of potential representation coefficients so that more bit-rate can be saved in the entropy coding\cite{hu2021learning}.\par

Ball{\'{e}} proposed a transformation framework consisting of multiple cascades of convolution layer and generalized divisive normalization (GDN)\cite{balle2015density,balle2017end}, which has excellent performance in image compression. Since then, there have been many transformation networks inspired by these frameworks. 

However, these approaches either increase complexity, such as nonlocal attention (NLA)\cite{chengtong2021tip,cheng2020learned}, invertible neural networks (INNs)\cite{xie2021enhanced}, augmented normalizing flows (ANFs)\cite{ho2021anfic} and wavelet-like transforms\cite{ma2020end}, to design more complex transformation modules, or introduce a priori hypothesis, for example, GSDN\cite{qian2020learning} assumes that the latent  obey a non-zero mean Gaussian distribution.

In this paper, we develop a new viewpoint, inspired by research about deep learning for the communication system, to interpret end-to-end image compression as a communication system\cite{shea2017DLPL}, which seeks to guide the design of transformation networks. Specifically, as shown in Figure \ref{ntc framework}, based upon the consistency in structures and optimization objectives in the two systems, the analysis transform,  quantization and  the synthesis transformation in image compression correspond successively to the transmitter, the channel and the receiver in communication system. 

Moreover, with the guidance of the communication system, we propose a novel transformation method based on signal modulation (TSM). We discovered that the previous transformation methods in image compression can be unified mathematically by TSM, and classified as the linear modulation techniques, which verifies the correctness of the guidance. Additionally, drawing on nonlinear modulation, three transform methods, transformation based on phase modulation (TPM), transformation based on frequency modulation (TFM), and transformation based on joint modulation (TJM), are proposed. \par
Lastly, in order to verify the effectiveness of TSM, the specific network architecture of the above methods and its residual blocks (ResTSM) are implemented. Our main contributions are summarized as follows:
\begin{itemize}
    \item We model the learned image compression framework as a communication system according to the consistency in structures and optimization objectives, which guides the design of the transformation.
    \item We propose a transformation method based on modulation techniques (TSM), which can be generalized to the previous transformation as the linear modulation techniques. Three transformation methods (e.g. TPM, TFM, TJM) are obtained by extending to nonlinear modulation. 
    \item An effective network architecture is implemented for above transformations and its residual block (ResTSM), which achieves a comparable performance with the SOTA transformation method in image compression. 
\end{itemize}
Experiment results on various datasets and different backbone networks show that ours method outperforms the most existing transformations in learned image compression, which verifies the effectiveness and robustness of TSM. 

%-------------------------------------------------------------------------

\section{Related Works}
\vspace{-0.5em}
\label{sec:related work}

\subsection{Learned Image Compression}
\vspace{-0.5em}
\begin{figure*}[t]
\begin{center}
\includegraphics[width=16.5cm]{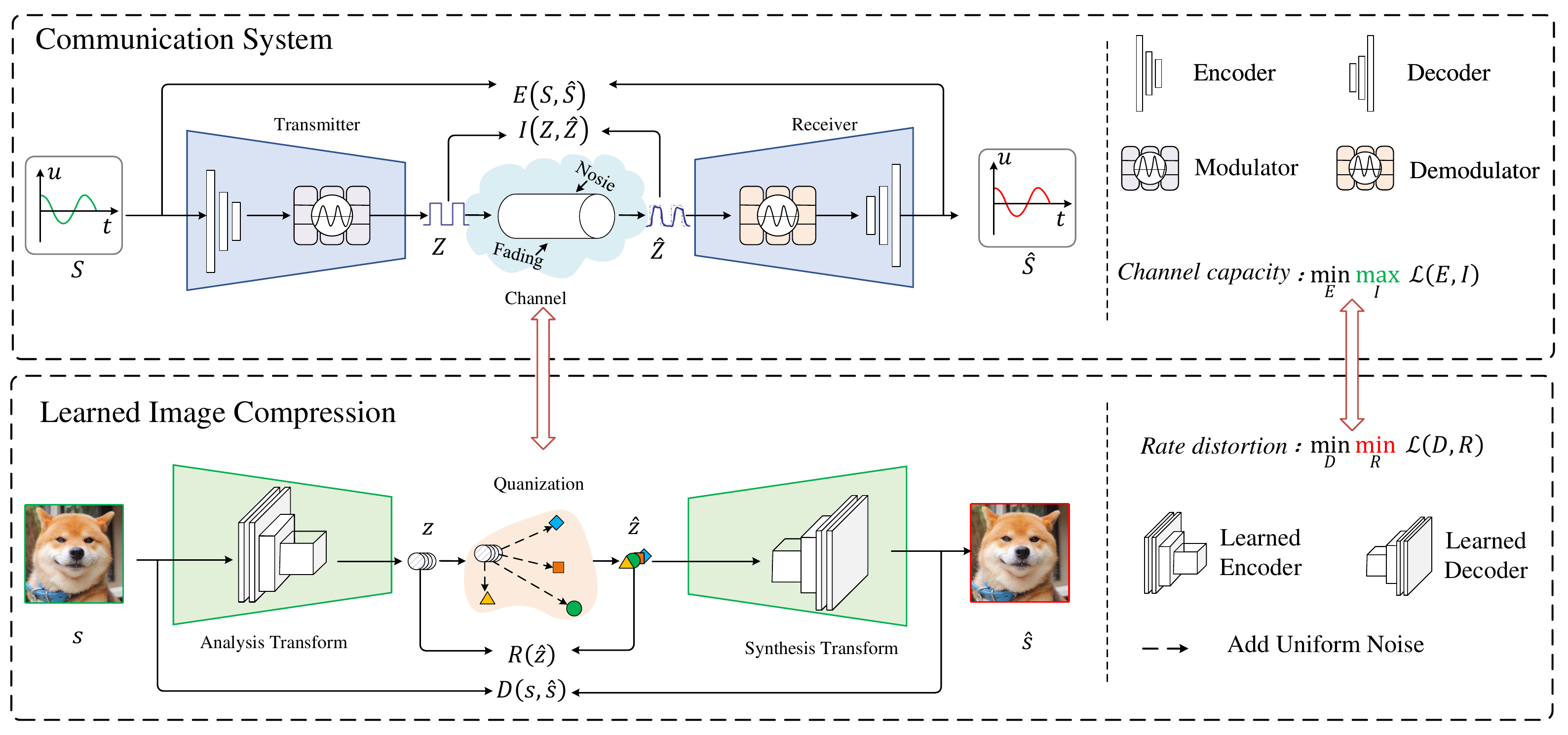}    %17
\end{center}
\vspace{-2em}
\caption{Illustration of the consistency between communication system and learned image compression. The top row represents communication system, where $\displaystyle S, Z, \hat{Z},\hat{S} $ denote the original signal, the transmitted signal, the received signal and the  estimate signal respectively, and $\displaystyle E$ represents the distortion between  $\displaystyle S$ and $\displaystyle \hat{S}$. $\displaystyle I$ represents the mutual information between $\displaystyle Z$ and  $\hat{Z}$. The bottom row represents communication system, where $\displaystyle s, z, \hat{z},\hat{s} $ denote the original image, the latent vertor, the discrete reconstructed vector and the restructured image respectively, and $\displaystyle D$ stands for the distortion between $\displaystyle s$ and $\displaystyle \hat{s}$. $\displaystyle R$ stands for the bitrate of $\displaystyle \hat{z}$. }
\vspace{-1em}
\label{ntc framework}
\end{figure*}

Learned image compression has made some significant breakthroughs and outperform classical methods\cite{balle2018variational,minnen2018joint,cheng2019learning,qian2020learning,xie2021enhanced,Cui_2021_CVPR,Gao_2021_ICCV}.

Most recent frameworks are VAEs based transform coding\cite{balle2020nonlinear}, which consists of transformation, quantization and entropy coding. The transformation is used to generate the compact latent representation and restore the reconstructed image.
End-to-end training requires replacing non-differentiable quantization with differentiable proxies such as additive noise\cite{balle2016endpcs,balle2017end} and soft assignments\cite{Guo2021SoftTH,agustsson2017soft}. 
The role of entropy coding is to encode the quantized the latent into a binary code stream, which benefits from the accurate entropy estimation\cite{cheng2020learned}, such as hyperpriors\cite{balle2018variational,hu2020coarse} and contextual model\cite{minnen2018joint}.

Ball{\'{e}} first proposed a practical transformation framework consisting of multiple cascades of convolution layer and GDN that gaussianize the local joint statistics\cite{balle2015density,balle2017end}, and is proved to be more efficient than other popular activation functions even the classical discrete cosine transform (DCT) \cite{balle18efficient}. 
Following methods in transformation can be roughly divided into two categories,including replacing GDN with other nonlinear transformations and modifying network structures. 
The former strategies focus on enhancing the GDN module, such as GSDN\cite{qian2020learning}, residual GDN (ResGDN)\cite{akbari2020learned} , shrinkage activation (SA)\cite{ogun2020shrinkage}, nonlocal attention (NLA)\cite{cheng2020learned,chengtong2021tip}.
The latter design the encoder as a whole for greater model expressiveness, for example, invertible neural networks (INNs)\cite{xie2021enhanced} for mitigating the information loss, augmented normalizing flows (ANF)\cite{ho2021anfic} for stacking multiple VAEs, and wavelet-like transforms\cite{ma2020end} for lossless transformation.\par
However, these approaches either increase the complexity of transformation or introduce a priori hypothesis that lacks an analysis of the overall framework.

\vspace{-0.5em}
\subsection{Autoencoder for End-to-End Communications Systems}
\vspace{-0.5em}

The concept of interpreting an end-to-end communications system as an autoencoder was first introduced in \cite{shea2017DLPL}. From a deep learning point of view, communication system can be seen as a particular type of autoencoder, where the transmitter and receiver in communication system correspond to the encoder and decoder in the VAE. The follow-up work extends this idea, using autoencoder to guide the design of the end-to-end communication system\cite{xiao2020data,hao2020deepgan,wu2019cnn,deniz2019machine}. In \cite{OFDM}, a conventional orthogonal frequency division multiplexing (OFDM) is proposed based on autoencoders. In \cite{DLSCC}, this idea is extended to semantic level, a deep learning based semantic communication system is proposed. \par
Such a setup creates a link between deep learning and communication systems and significantly improves the performance of the communication system\cite{OFDM}. However, whether this idea can be further extended by considering VAE as a special communication system and using the communication system to guide the design of VAE-based image compression is a work worth investigating.

%-------------------------------------------------------------------------

\section{Communication System for Learned Image Compression}
\label{sec:method}
In this section, we will illustrate the mapping relationship between learned image compression systems and communication systems. With this mapping relationship, the transformation in the compression system is interpreted as the modulation module. Subsequently, a transformation method based signal modulation (TSM) are proposed.

\subsection{Consistency Analysis}

Contrary to the previous works, the VAE-based learned image compression framework is considered as a special communication system from communication perspective. As illustrated in Figure \ref{ntc framework}, the transceiver of communication system is consistent with learned image compression framework in structure and optimization objectives. \par

\noindent \textbf{Consistent Structure}
% \hangindent 1em
% \hangafter=0
% \indent
\ A communication system, in its simplest form, consists of a transmitter, a channel, and a receiver, as shown in Figure \ref{ntc framework}.  The transmitter applies the transformation to $\displaystyle S$ to generate  $\displaystyle Z \in \mathbb{R}^{n}$. $\displaystyle Z$ generates $\hat{Z} \in \mathbb{R}^{n}$ through the channel. After receiving $\hat{Z}$, the receiver applies the transformation to produce $\hat{S}$. 
Typically, the simplest learned image compression framework is based on auto-encoder, and consists of three parts: the analysis transform $\displaystyle g_{a}(s)$, quantization $\displaystyle q$ , and the synthesis transform $\displaystyle g_{s}(\hat{y})$.
An image $\displaystyle s$ is transfomed to $\displaystyle z$ by $\displaystyle g_{a}(s)$. $\displaystyle z$ is subjected to quantization $\displaystyle q$, yielding $\hat{z}$ that can be compressed to bit stream by standard arithmetic coding algorithm.  Conversely, $\displaystyle g_s$ converts  $\displaystyle \hat{z}$ into $\displaystyle \hat{s}$. 
Comparing the two above systems, there is a consistent relationship between their structures, where the analysis transform, quantization, and the synthesis transform correspond successively to transmitter, channel and receiver.

\noindent \textbf{Consistent Optimization Objectives}
% \hangindent 1em
% \hangafter=0
% \noindent
The optimization objectives of the two systems are described in Figure \ref{ntc framework}. 
The optimization goal of the communication system, denoted as ${\mathop{\min} \limits_E \mathop{\max} \limits_I  {\mathcal{L}(E,I)}}$, is to minimize the distortion of the message, metrics like signal-to-noise (SNR) ratio or bit error rate (BER) are usually used, while maximizing the capacity of communication system, given by mutual information $\displaystyle I$ between $\displaystyle Z$ and $\hat{Z}$.
% where $\displaystyle E$ represents the distortion between the original message $\displaystyle S$ and the estimate message $\displaystyle \hat{S}$ where metrics like signal-to-noise(SNR) ratio or bit error rate(BER) are usually used. $\displaystyle I$ represent the capacity or the data transmission rate of a communication system, which is given by mutual information between transmitted signal $\displaystyle Z$ and received signal $\hat{Z}$.\par
In image compression, the goal of image compression, denoted as $ \mathop{\min} \limits_D \mathop{\min} \limits_R  {\mathcal{L}(D,R)}$, is to minimize the distortion of the image, such as mean-squared error or multi-scale structural similarity, while minimize the bit rate of latent features, obtained by the entropy of  $\hat{z}$. 
% Where $\displaystyle D$ stands for the distortion, such as mean-squared error or multi-scale structural similarity, $\displaystyle R$ stands for the bitrate of the quantized latent features.
Comparing the optimization terms of the two systems, both $\displaystyle E$ and $\displaystyle D$ represent the distance between the original signal and the reconstructed signal, so minimizing $\displaystyle E$ and minimizing $\displaystyle D$ are identical. In information theory, maximizing $\displaystyle I$ and minimizing $\displaystyle R$ is a dual problem (the proof is in the supplementary materials), so the two terms are equivalent. This means that the optimization objectives of the two systems are consistent.

\begin{figure}[t]
\begin{center}
\includegraphics[width=8cm]{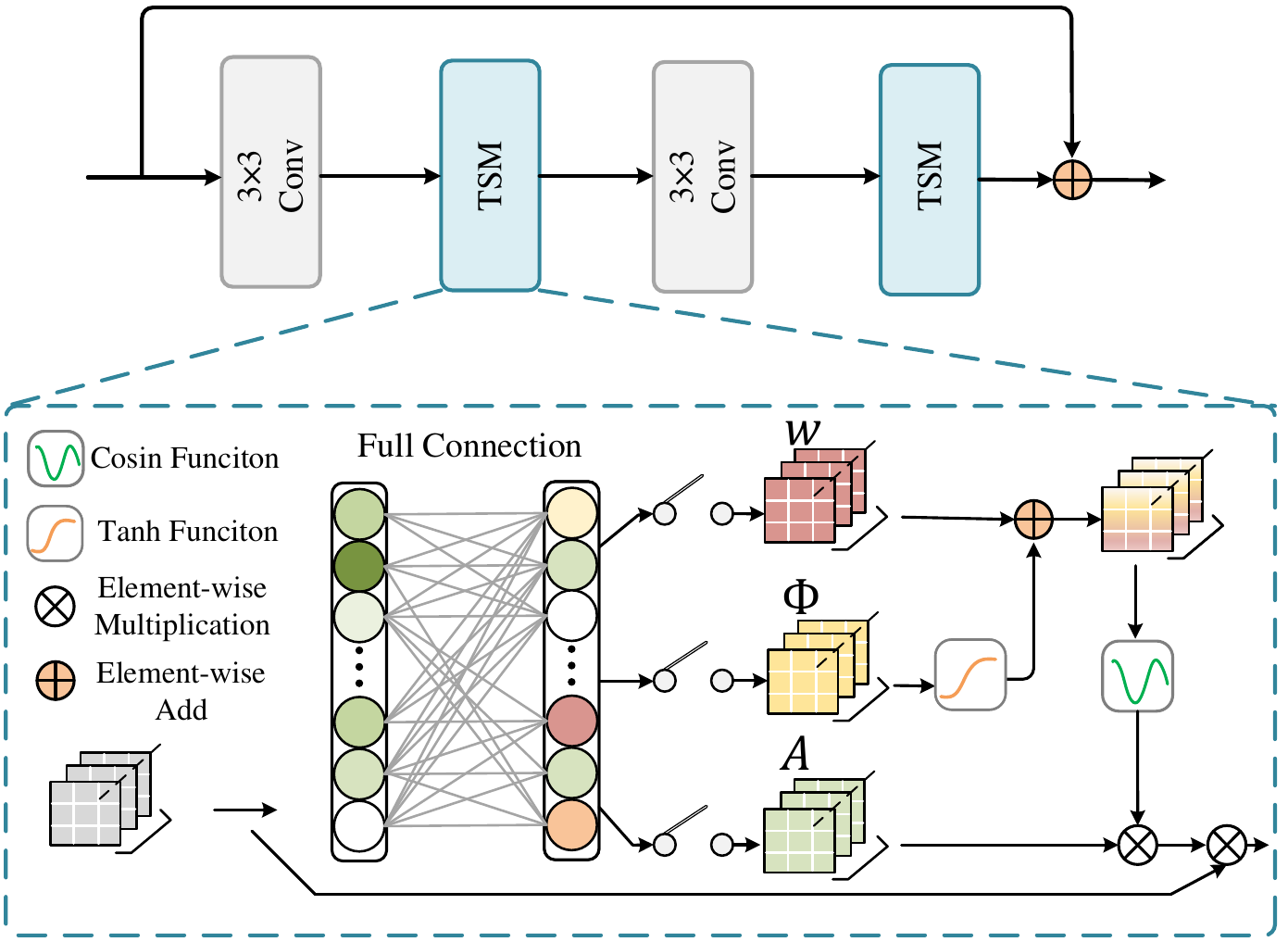}
\end{center}
\vspace{-1.5em}
\caption{Illustration of the proposed transforms based on Signal Modlation (TSM) and its residual block (ResTSM). The switch means that three branches $A,\ \phi \ \text{and} \ \omega$ can be controlled to obtain different transformation methods: when only $\displaystyle A$ or $\phi$ or $\omega$ branch is available, this corresponds to TAM, TPM and TFM respectively.}
\label{TSM}
\vspace{-1.5em}
\end{figure}

\noindent Based on the above analysis, the image compression framework can be modeled as a communication system according to the consistency in structures and optimization goals, so the technology of communication systems can be applied to guide the design of modules in image compression.

\subsection{Transform based on Signal Modulation}
\vspace{-0.5em}

Modulation is an indispensable module in transmitter of communication systems, which is used to shift the frequency spectrum of the signal. 
The modulation of a signal $\displaystyle x \to f(x)$ proceeds as follows:
\begin{align}
    % f(x) &= x \cdot c(x)  \\  \notag
    % \text{s.t.  }   c(x) &= A(x) \cdot cos(\omega(x) + \phi(x))
    f(x) = x \odot \underbrace{A(x)cos(\omega(x) + \phi(x))}_{C}
\vspace{-0.5em}
\label{modulation equation}
\end{align}
\begin{table}[b]
\vspace{-1em}
\caption{$\displaystyle f(x)$ in Equation.\ref{modulation equation} is generalized to various existing nonlinear transformations when the $A, \omega, \phi$ of carrier take different values. $\displaystyle \theta$ in SA\cite{ogun2020shrinkage} means a learnable parameter,  Attention weights $\displaystyle W(x)$ in NAL\cite{cheng2020learned} calculated by $\displaystyle sigmoid$ or $\displaystyle softmax$ function. }
\vspace{-0.5em}
\label{table:nonlieartrans}
\centering
\resizebox{\linewidth}{!}
{
\begin{tabular}{cccc}
\toprule
\textbf{Method} &$\displaystyle A(\cdot)$ &$\displaystyle \omega(\cdot)$ &$\displaystyle \phi(\cdot)$  \\ \hline \hline
\makecell[c]{ReLU\cite{hindon2010relu}} &$
\begin{cases}
0& x \le 0 \\
1& x > 0 
\end{cases}$ &\multicolumn{2}{c}{$\displaystyle cos(\omega(x) + \phi(x)) \equiv{1} $}         \\ 
\makecell[c]{GDN\cite{balle2015density}}       &${{{1}} \over {\mathop {({\beta ^2}  + {\sum _j}{\gamma _j}x_j^2)}\nolimits^{1 \over 2}}}$     & \multicolumn{2}{c}{$\displaystyle cos(\omega(x) + \phi(x)) \equiv{1} $}      $ $    \\  
\makecell[c]{SA\cite{ogun2020shrinkage}}  & $
\begin{cases}
0&  (\frac{x}{\theta}) \in {[\footnotesize{-0.5,0.5}]}  \\
1& \text{otherwise}
\end{cases}$    & \multicolumn{2}{c}{$\displaystyle cos(\omega(x) + \phi(x)) \equiv{1} $}       \\ 
\makecell[c]{NAL\cite{cheng2020learned}}   &$W(x) \in  [0,1]$     & \multicolumn{2}{c}{$\displaystyle cos(\omega(x) + \phi(x)) \equiv{1} $}    \\  \hline
\makecell[c]{TPM}   & 1     & \makecell[c]{$\displaystyle k\pi$ }  &$\displaystyle \phi(x)$   \\ 
\makecell[c]{TFM}   & 1    &$\displaystyle \omega(x)$  &$\displaystyle k\pi$  \\ 
\makecell[c]{TJM}   & $\displaystyle A(x)$    &$\displaystyle \omega(x)$  &$\displaystyle \phi(x)$   \\ 
\bottomrule
\end{tabular}
}
\label{table generalized}
\end{table}
where $\displaystyle x$ denotes input signal, $\displaystyle C$ is carrier signal, $\displaystyle f(x)$ denotes modulated signal, 
$A, \omega, \phi$  represent, in turn, amplitude, frequency and phase of the carrier, $\odot$ represents the element-wise multiplication.
In learned image compression, the transform $g_a(x)$ consists of a linear decomposition followed by a joint nonlinear transform.

As illustrated in Table \ref{table generalized}, previous nonlinear transform methods in learned image compression can be generalized by Equation (\ref{modulation equation}).

% \vspace{0.5em}
\begin{figure}[H]
\begin{center}
\includegraphics[width=8.5cm,height=5.7cm]{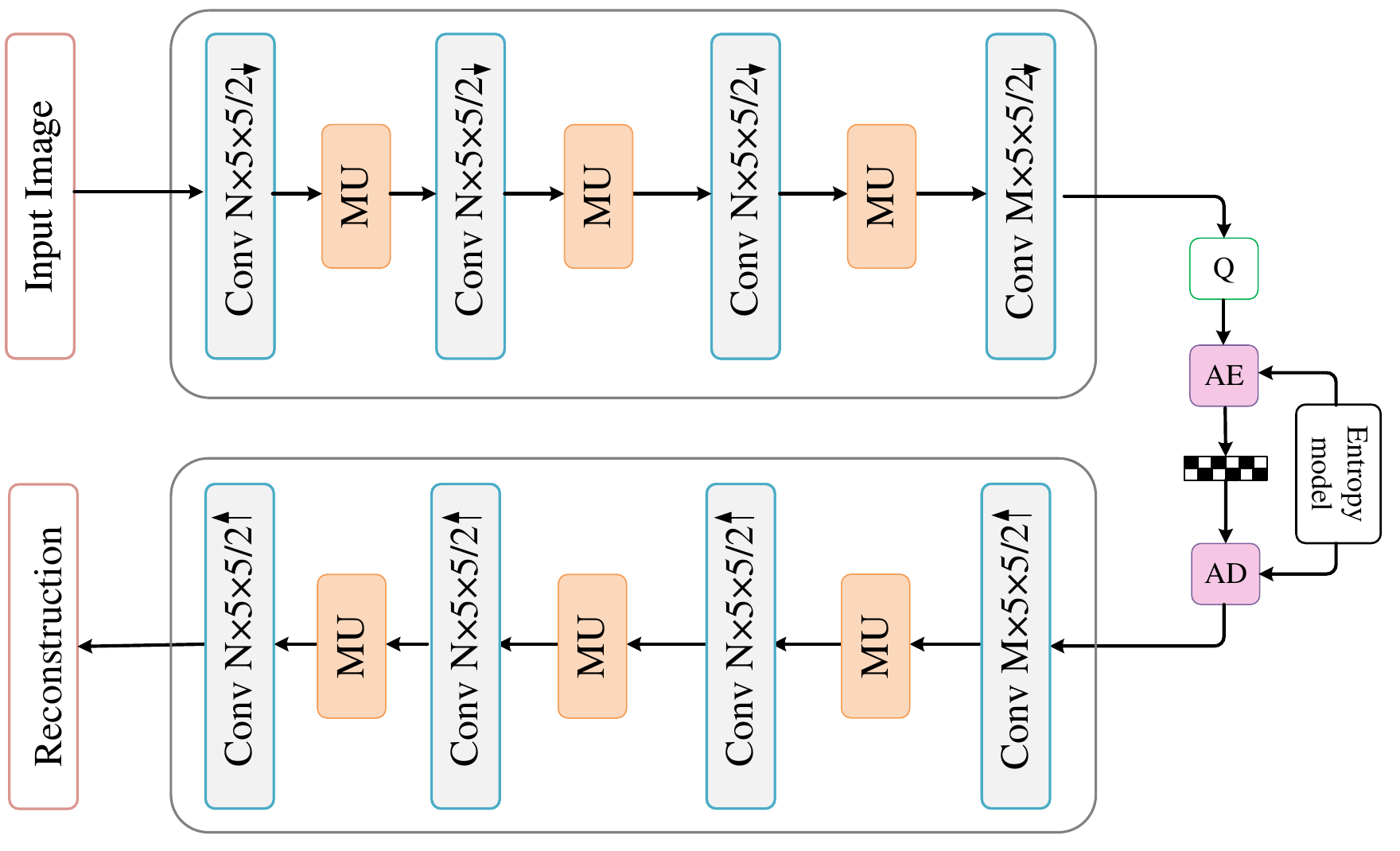}
\end{center}
\vspace{-1.5em}
\caption{Illustration of our methods on ICLR2018\cite{balle2018variational} network architecture. MU is a collective term for our methods, including TPM, TFM, TJM and its residual block (ResTSM). Entropy model represents the hyperprior model. Q represents quantization, and AE, AD represent arithmetic encoder and arithmetic decoder, respectively. Convolution parameters are denoted as: number of filters $\times$ kernel support height $\times$ kernel support width / down- or upsampling stride, where ↑ indicates upsampling and ↓ downsampling.}
\label{TSM on balle2018}
\vspace{-1em}
\end{figure}
The above analysis shows that the current nonlinear transformations in image compression are only varied in form of amplitude, corresponding to the signal linear modulation technique, which limits the nonlinear capability of the transformations. We can achieve more nonlinear transformations by changing the frequency and phase corresponding to the nonlinear modulation technique.\par
\begin{figure*}[t]
\begin{center}
\includegraphics[width=17cm]{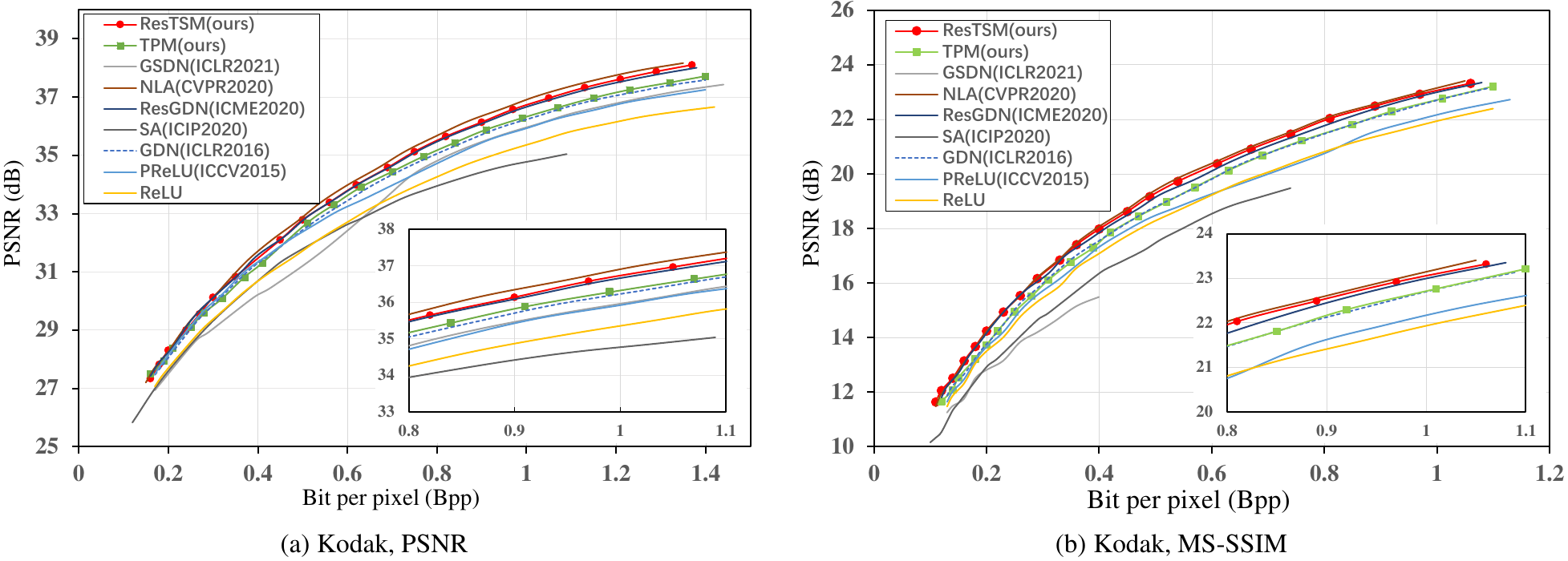}
\includegraphics[width=17cm]{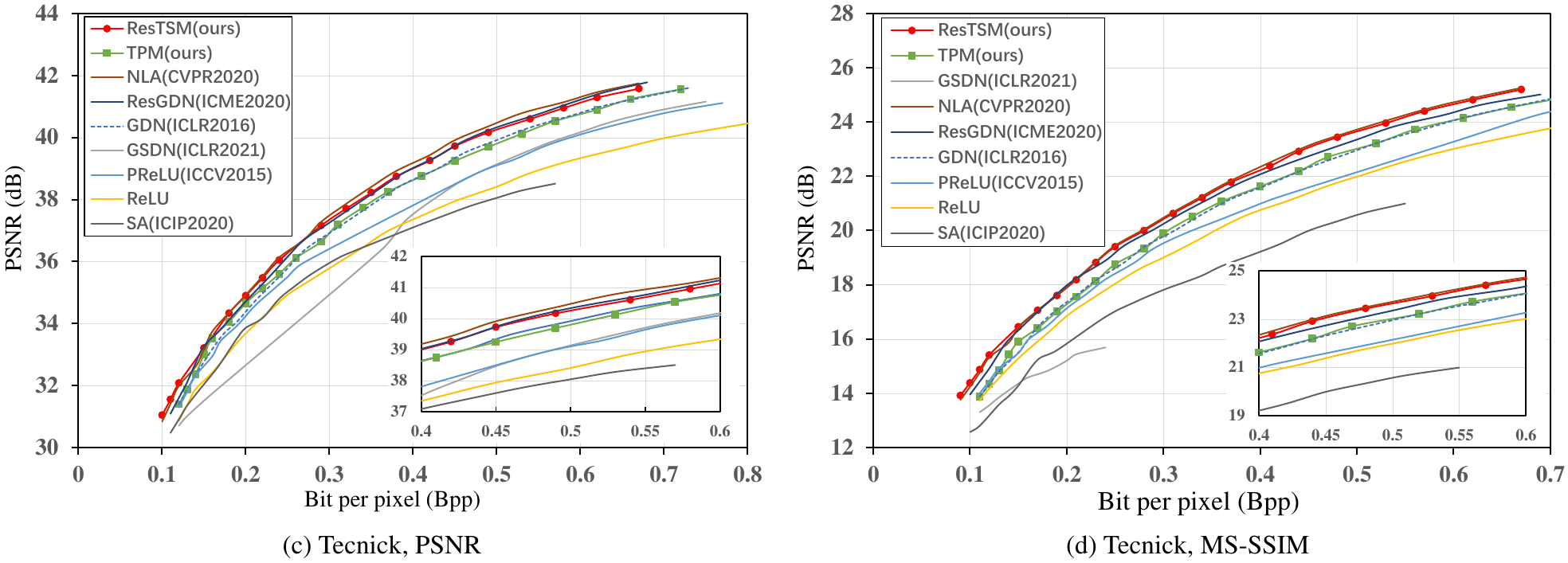}
\end{center}
\vspace{-2em}
\caption{Rate-Distortion Performance of our proposed transformation and the other methods on the Kodak and Tecnick datasets. The basic model is CVPR2021\cite{Cui_2021_CVPR}, and all experiments only  operate on the nonlinear transformation without changing the whole neural network architecture and hyperparameters. PSNR and MS-SSIM are used as the distortion metrics. The MS-SSIM values are converted into decibels $\displaystyle -10log_{10}(1-d)$ where $\displaystyle d$ refers to the MS-SSIM value, for a clear illustration. The anchor for calculating the BD-rate gain is CVPR2021\cite{Cui_2021_CVPR} with GDN while distortion is measured by PSNR.}
\vspace{-0.5em}
\label{RDperformance}
\end{figure*}

\noindent \textbf{Transform based Nonlinear Modulation}
Other transformations based modulation method can be obtained obviously by choosing the values of $A, \omega, \phi$. 
% Choosing $\displaystyle A(x)=a, \omega(x)=b$, where $\displaystyle a$ and $\displaystyle  b$ both are constants. 
For simplicity of representation and implementation, we introduce two simple nonlinear modulation techniques: Frequency Modulation (FM) and Phase Modulation (PM). As shown in Table. \ref{table generalized}, when $\displaystyle A=1, \omega=k\pi, k$ is an integer, the phase of the output signal gets shifted depending upon the input, which corresponds to phase modulation (TPM). Besides, when $\displaystyle A=1, \phi=k\pi, k$ is an integer, the frequency of the output signal only varies with the input, which corresponds to frequency modulation (TFM). Moreover, varying Amplitude, Phase and Frequency at the same time corresponds to joint modulation (TJM). 

It is notable that the inversion of the TSM at the decoder is formally identical to the TSM due to the characteristics of the trigonometric function, which is an advantage over the GDN.\par

In this subsection, we propose a transformation based on signal modulation (TSM) that can generalize the previous works. Then three  transformation methods, including TPM, TFM and TJM, are proposed.  Next, we will implement the above methods in detail and describe the existing transformation network according to modulation techniques.
\begin{table}[H]
\vspace{1em}
\caption{Comparison of the complexity of different nonlinear transformation layer. For a fair comparison, we uniformly use the first layer of nonlinear transformation parameters in the CVPR2021\cite{Cui_2021_CVPR}, where the number of input and output channels is 192, feature size is $256 \times 256$.
The anchor for calculating the BD-rate gain is CVPR2021\cite{Cui_2021_CVPR} with GDN. The $\textcolor{red}{\blacktriangle}$ denotes improvement, and $\textcolor{green}{\blacktriangledown}$ denotes deterioration.}
\vspace{-0.5em}
\label{table:computational_cost}
\centering
\renewcommand\arraystretch{1.8}
\resizebox{\linewidth}{!}{
\begin{tabular}{ccccc}
\toprule
\textbf{Method} &\makecell[c]{BD-Rate\\Kodak}$\uparrow$ &\makecell[c]{BD-Rate\\Tecnick}$\uparrow$ &{GFLOPs}$\downarrow$  &\makecell[c]{Para.\\ $\times  10^{4}$}$\downarrow$  \\ \hline \hline
GDN\cite{balle2015density}  &Anchor &Anchor &0.61 &3.71 \\ \hline
GSDN\cite{qian2020learning}  &-14.11\%\scriptsize${\textcolor{green}{\blacktriangledown}}$ &-26.20\% \scriptsize${\textcolor{green}{\blacktriangledown}}$ &1.22\scriptsize${\textcolor{green}{\blacktriangledown {100}\%}}$ &7.39\scriptsize${\textcolor{green}{\blacktriangledown {100}\%}}$ \\
ResGDN\cite{akbari2020learned}  &6.06\%\scriptsize${\textcolor{red}{\blacktriangle}}$   &6.03\% \scriptsize${\textcolor{red}{\blacktriangle}}$ &12.08\scriptsize${\textcolor{green}{\blacktriangledown {1880}\%}}$  &73.75\scriptsize${\textcolor{green}{\blacktriangledown {1880}\%}}$  \\
NAL\cite{cheng2020learned}  &8.59\%\scriptsize${\textcolor{red}{\blacktriangle}}$  &8.77\% \scriptsize${\textcolor{red}{\blacktriangle}}$ &42.89\scriptsize${\textcolor{green}{\blacktriangledown {6931}\%}}$  &261.75\scriptsize${\textcolor{green}{\blacktriangledown {6955}\%}}$  \\  \hline
TPM (ours)  &1.60\%\scriptsize${\textcolor{red}{\blacktriangle}}$  &1.11\% \scriptsize${\textcolor{red}{\blacktriangle}}$ &0.61\scriptsize${\textcolor{red}{- {0}\%}}$ &3.71\scriptsize${\textcolor{red}{- {0}\%}}$   \\
ResTSM(ours)  &6.13\%\scriptsize${\textcolor{red}{\blacktriangle}}$  &7.53\%\scriptsize${\textcolor{red}{\blacktriangle}}$
&12.69 \scriptsize${\textcolor{green}{\blacktriangledown {1980}\%}}$  &77.43\scriptsize${\textcolor{green}{\blacktriangledown {1987}\%}}$  \\
\bottomrule
\end{tabular}
}
\label{complexity}
\end{table}
\begin{figure*}[t]
\begin{center}
\includegraphics[width=17.5cm]{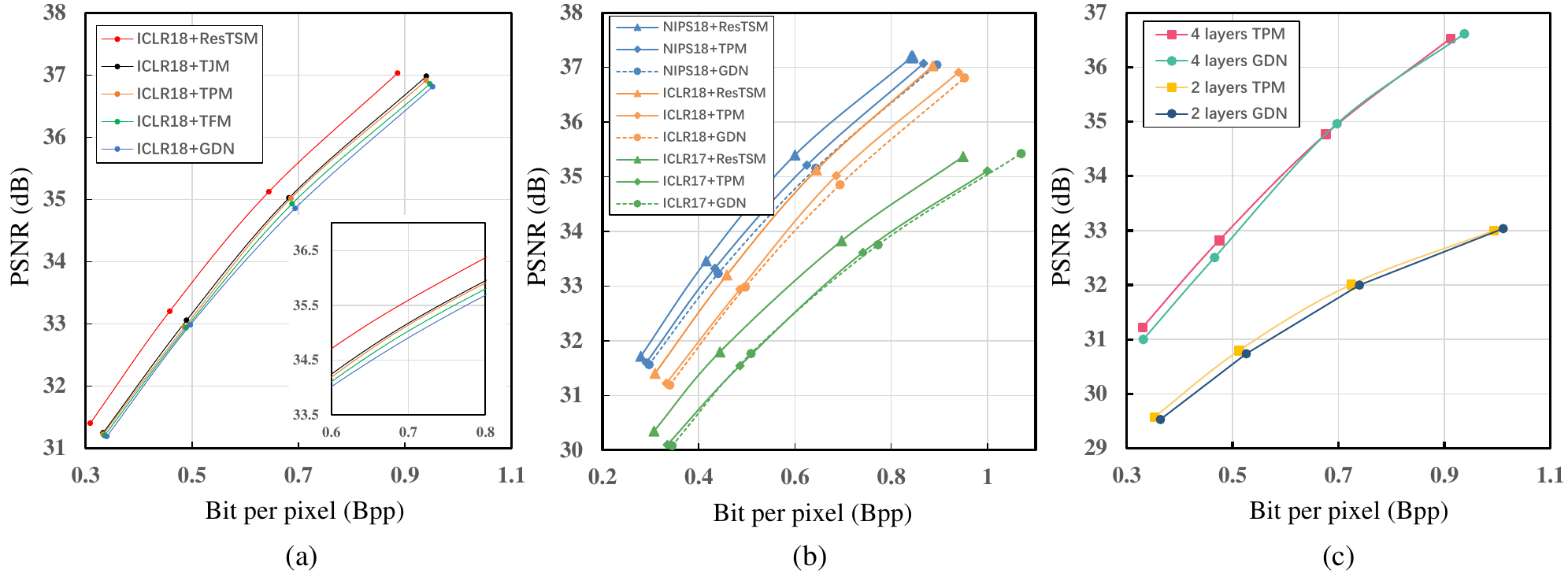}
\end{center}
\vspace{-2em}
\caption{(a) Comparison of R-D performance of ICLR2018\cite{balle2018variational} with different modulation techniques on the Kodak dataset. (b) Comparison of R-D performance of our methods on different VAE-based image compression architecture on the Kodak dataset. (c) Comparison of R-D performance on different network depth by modifying the number of cascading layers on the basic model ICLR2018\cite{balle2018variational}. 2 layers GDN/TPM means there are 2 layers of cascaded GDN/TPM and down-sampling convolution in network architecture.}
\label{ablation study}
\vspace{-1em}
\end{figure*}
% -----------下一节--------------------
% 3. 引入相干解调, 正交调制等技术
% 4. 解调的过程是如何逆变换的;

\subsection{Network Architecture}
% We first implement the basic module of NTSM with CNNs. Then we apply the residual structure for increasing the nonlinear capability of NTSM. Finally, we propose a codec parameter sharing structure using coherent demodulation techniques to reduce the number of parameters.\par
The key to implementing the TSM is to create a mapping from  $\displaystyle x$ to $\displaystyle A$, $\displaystyle \omega$  and $\displaystyle \phi$ of the carrier signal, which can be achieved by ANNs. As demonstrated as the Figure \ref{TSM}, the mapping is constructed by three fully connected layers (FCs), where different architectures are obtained by controlling whether  $\displaystyle A$, $\displaystyle \omega$  and $\displaystyle \phi$
are involved in the calculation of Equation (\ref{modulation equation}), and $\displaystyle tanh$ function is used to constrain the phase. 
Previous work has demonstrated that more nonlinear transform structure can increase the compression performance of the network\cite{akbari2020learned}. Meanwhile, in order to obtain deeper learning of image statistics and faster convergence, the concept of identity shortcut connection is introduced to basic module which denotes as ResTSM, as shown in Figure \ref{TSM}.  \par
Figure \ref{TSM on balle2018} demonstrates the proposed architecture derived from the ICLR2018 model\cite{balle2018variational}, which can be interpreted as mutil-stage modulation process in communication systems. The analysis transform $\displaystyle x \to g_a(x)$ can be representd as:
\begin{equation}
\begin{aligned}
    g_a(x) =& {M_k}({M_{k - 1}} \cdots {M_0}(x) \cdots ) \\
{M_k}(x)=&
\begin{cases}
{f[{H_ \downarrow }(x)]}& \text{k} \in \{ {0,1,...,n-1} \}  \\
{{H_ \downarrow }(x)}&  \text{k} = \{ {n} \} \\
\end{cases}
\end{aligned}
\end{equation}
where $\displaystyle n$ represents number of cascade stages, $\displaystyle {M_k}$ is modulation unit of the $\displaystyle k$-th stage, $H_ \downarrow $ denotes down-sampling convolution, and $f(\cdot)$  is nonlinear transformation calculated by process of Equation (\ref{modulation equation}). The synthesis transform $\displaystyle \hat{z} \to g_s(\hat{z})$ is the opposite of the above process, which is mutil-stage demodulation process: 
\begin{equation}
\begin{aligned}
    g_s(\hat{z}) =& {M^{-1}_k}({M^{-1}_{k - 1}} \cdots {M^{-1}_0}(z) \cdots ) \\
{M^{-1}_k}(\hat{z})=&
\begin{cases}
{{H_ \uparrow }(\hat{z})}&  \text{k} = \{ {0} \} \\
{f^{-1}[{H_ \uparrow }(\hat{z})]}& \text{k} \in \{ {1,2,...,n} \}  \\
\end{cases}
\end{aligned}
\end{equation}
where $\displaystyle {M^{-1}_k}$ is demodulation unit of the $\displaystyle k$-th stage, $H_ \uparrow $ denotes up-sampling convolution, and $f^{-1}(\cdot)$  is inverse of nonlinear transformation.
Specifically, in the ICLR2018\cite{balle2018variational} model, $\displaystyle n$ is taken as 3, $f(\cdot)$ is GDN and $f^{-1}(\cdot)$  is IGDN. In order to verify the validity, we substitute all the GDN and IGDN layers in the ICLR2018\cite{balle2018variational} model with the proposed TSM. \par

\begin{figure*}[t]
\begin{center}
\includegraphics[width=17cm]{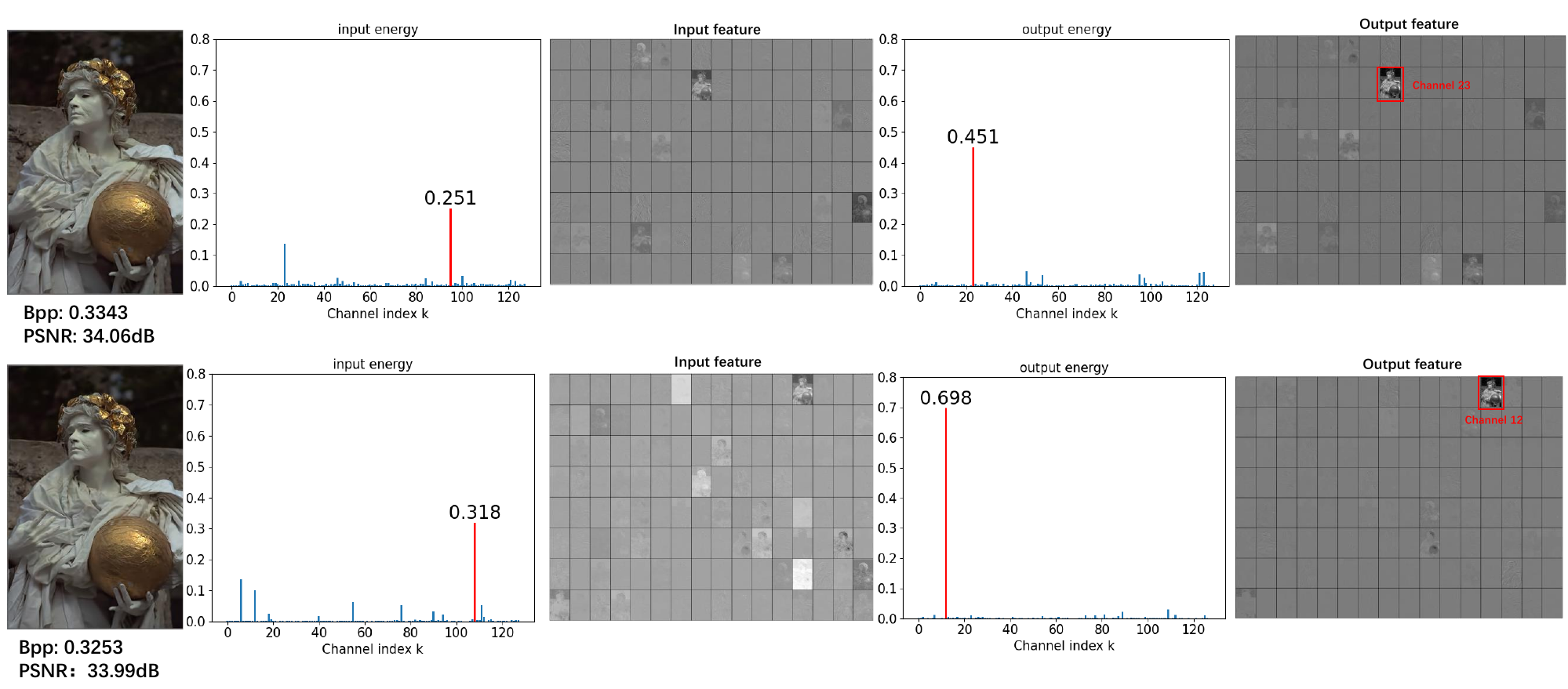}
\end{center}
\vspace{-2em}
\caption{Visualization for different transformation methods using kodim17. The top row indicates GDN method, and the bottom row is TPM. From left to right column indicate the reconstructed image, the energy distribution of the channel before the transformation, and the energy distribution after the transformation, respectively.}
\label{energy distribution}
\vspace{1em}
\end{figure*}

\section{Experiments}
% In this section, we first introduce the implementation details of the proposed method, and then fully evaluate and analyze the performance of our method.

\subsection{Implementation Details}
% 附录放一张框架图
We evaluate the proposed transformation methods, including TFM, TPM, and TJM, based on four previous architectures: ICLR2017\cite{balle2017end}, ICLR2018\cite{balle2018variational}, NIPS2018\cite{minnen2018joint} and CVPR2021\cite{Cui_2021_CVPR}. The former three architectures, the mainstream VAE-based image compression architectures, are used for ablation experiments. The last one, the state-of-the-art variable-rate image compression framework, is for performance analysis.
As shown in Figure \ref{TSM on balle2018} using ICLR2018 as an example, we do not change any other architectures except replacing the GDN layer with the proposed methods to ensure fair comparisons.\par
\noindent \textbf{Datesets and data processing} We train our compression model on CLIC training dataset\cite{CLIC2020} which is randomly cropped into 247576 images of $256 \times 256$ pixels. For comparison, we test our model on the standard kodak dataset with 24 images of $512 \times 768$ or $768 \times 512$ pixels and the tecnick dataset with 40 images of $1200 \times 1200$ pixels. We evaluate the rate distortion performance of our model and baseline models on bit-per-pixel (bpp), peak signal-to-noise ratio (PSNR) and multi-scale structural similarity (MS-SSIM) metric.

\noindent \textbf{Training details} The baseline models and the proposed model are implemented based on the open-source library \cite{begaint2020compressai} in PyTorch. We implement the above architectures, and for each baseline, we train under the same settings. The detailed experimental settings are as follow:\par
All models (\textbf{ICLR2017, ICLR2018, NIPS2018, CVPR2021}) are trained for 100 epochs with a batch size of 16 using Adam optimizer, and the learning rate is set to $10^{-4}$ for the first 64 epochs and reduced to half for the last 36 epochs. Lagrange multiplier $\lambda$ controls the trade-off between bit rate and distortion. Channel number $M, N$ are set to $192, 128$ respectively for low bit-rate models when $\lambda \in \{0.0018, 0.0035, 0.0067, 0.0130\}$ and $320, 192$ for high bit-rate models when $\lambda \in \{0.0250, 0.0483, 0.0932, 0.1800\}$ separately. The threshold of gradient clipping is 1 to stable training.

\subsection{Rate-distortion Performance}
% 6张图
In order to show the advantage of our methods, we first evaluated our model by obtaining the average rate-distortion performance in terms of PSNR and MS-SSIM. We further measured the reductions in BD-rate of our methods, which is defined as the
average saving in bitrate between two models for a given quality metric. And we compare it with a series of variants of nonlinear transformations, which mainly repalce GDN with advanced modules including ResGDN\cite{akbari2020learned}, GSDN\cite{qian2020learning}, Attention\cite{cheng2020learned,chengtong2021tip}, shrinkage activate\cite{ogun2020shrinkage}, PReLU\cite{He2015DelvingDI}, ReLU\cite{hindon2010relu}. \par
The R-D performance on Kodak dataset is illustrated in Figure \ref{RDperformance}, TPM achieves a 1.60$\%$ reduction in BD-rate than GDN and ResTSM obtains comparable R-D performance with that of attention methods and even better R-D performance than ResGDN methods. Regarding MS-SSIM, TPM also achieves a same performance in BD-rate than GDN. It should be noticed that the number of parameters and computation of TPM are the same as GDN, while the number of parameters and computation of ResTSM are comparable with ResGDN, which are 29.59\% of the attention module (as shown in Table \ref{table:computational_cost}). 
Moreover, the performance on high-resolution dataset, Tecnick dataset, show that TPM yields better coding performance than GDN and  ResTSM obtains comparable R-D performance with NAL and ResGDN in terms of both PSNR and MS-SSIM. It is notable that since the priori assumption of GSDN is that the image obeys a Gaussian distribution with non-zero mean, the performance of GSDN will be deteriorated when the entropy model is a zero-mean Gaussian model.

\begin{figure*}[t]
\begin{center}
\includegraphics[width=17cm]{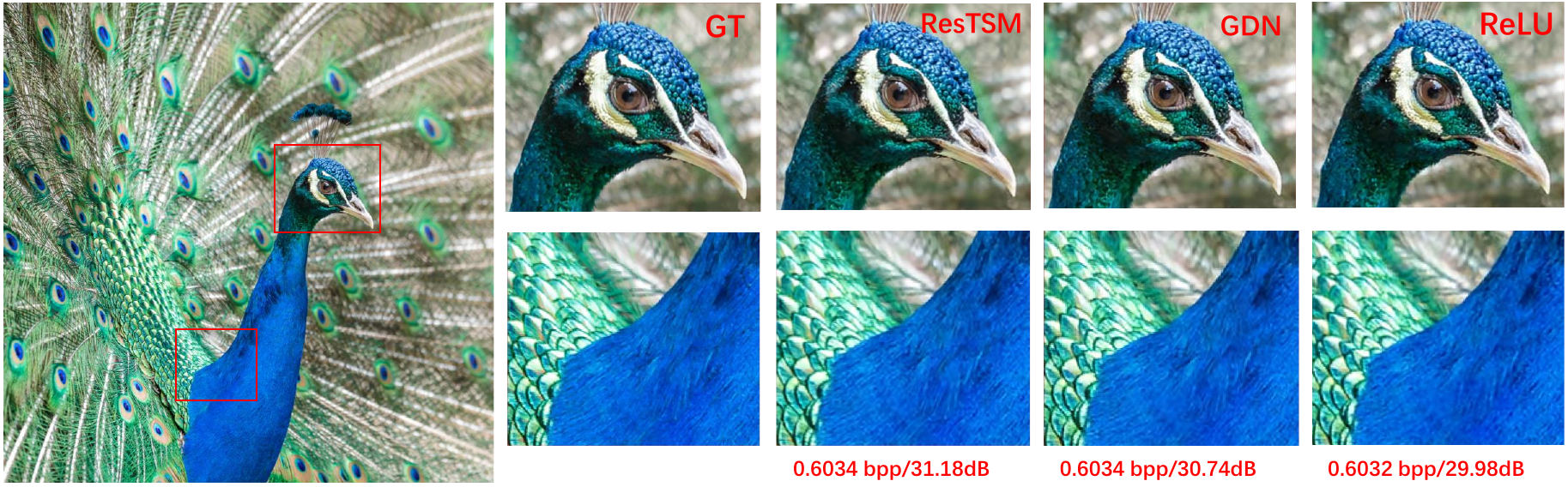}
\end{center}
\vspace{-2em}
\caption{Visualization comparion of reconstructed images with approximately 0.6 bpp. The red annotations in the cropped images indicate bpp/PSNR (dB) of an entire image.}
\label{subjective result}
% \vspace{-1em}
\end{figure*}

\subsection{Complexity Analysis}
As shown in Table \ref{table:computational_cost}, we compared the number of parameters, the amount of computation and the inference time for our proposed TPM and TSM methods  with other nonlinear transforms. Since the above comparison parameters need to depend on the specific network, we uniformly use the first layer of nonlinear transformation parameters in the ICLR2018\cite{balle2018variational}, where the number of input and output channels is 192, the feature size is $128 \times 128$. Obviously, 
The highest complexity module is attention module\cite{cheng2020learned}. Our light model TPM has the same FLOPs and parameters with GDN while achieving better R-D performance on different datasets and two evaluation metrics, which proves that our method is more efficient than GDN. And our best model can greatly obtain performance gain by increasing certain computational costs, which has more parameters than GDN and 29.59\% of the attention module.

\subsection{Qualitative Results Analysis}
As shown in Figure \ref{subjective result}, to demonstrate that our method can generate more visually pleasant results, we visualize some reconstructed images for qualitative performance comparison. The reconstructed images with approximately 0.60 bpp, which are generated from above nonlinear transform methods. We observe that our method optimized by MSE has clearer edges and textures than other methods. 
% More qualitative results are included in supplementary materials.

\section{Ablation Studies}
\vspace{-0.5em}
In order to verify the effectiveness of our proposed methods and understand the contribution of each of them, we first evaluated the impact of different modulation techniques and then
confirmed the generalization capability on multiple backbone network architectures by replacing the nonlinear transform layers. Ablation studies only operate on the nonlinearities without change of the whole neural network architecture and hyperparameters. In addition, to understand the role of modulation, we visualized the features before and after the transformation. The following can be summarized from the ablation results:

\textbf{TSM Methods} Figure \ref{ablation study}.(a) shows the results on ICLR2018\cite{balle2018variational} with different modulation techniques, including GDN, frequency modulation (TFM), phase modulation (TPM), joint modulation (TJM) and ResTSM. 
When parameters are equal in number, TFM has better results than GDN, yields a 1.42$\%$ reduction in BD-rate, while TPM can achieve the best result among them, yields a 2.76$\%$ reduction in BD-rate. Moreover, TJM and ResTSM can bring 3.66$\%$ and 11.50$\%$ reduction in BD-rate, respectively, which validates that the architectural design we made to the modulated based transform methods are effective.

\textbf{Visualization for Modulation}  
We show the energy distribution of different channels in ICLR2018\cite{balle2018variational} by using various nonlinear transformation, including GDN and NPM, in Figure \ref{energy distribution}, this visualization method is referred to Cheng2020\cite{cheng2020energy}. For illustration purposes, we define a channel energy ratio 
$\displaystyle {e_i} = {{{E_i}} \over {\sum\limits_i {{E_i}} }}$, where $\displaystyle E_i$ denotes the energy of channel index $\displaystyle i$. When the transformation are GDN, and TPM respectively, the maximum $\displaystyle e_i$ after the transformation are 0.451 and 0.698 in order. We can conclude that modulation techniques have better energy concentration performance than GDN, and compact energy contributes to compression performance\cite{cheng2020energy,cheng2019learning}, which is probably the reason why TPM is more effective. More visualization results are included in supplementary materials.

\textbf{Generalizability of TSM} 
We first verify the performance of our methods on different VAE-based image compression architectures, including ICLR2017\cite{balle2017end}, ICLR2018\cite{balle2018variational} and NIPS2018\cite{minnen2018joint}, by substituting all the GDN layers with the proposed NTM. As shown in Figure \ref{ablation study}.(b), TPM attained 0.77$\%$, 2.76$\%$ and 3.52$\%$ reduction than GDN in BD-rate on the three above networks in order, and the best method ResTSM achieved  12.41$\%$, 11.50$\%$ and 12.41$\%$ reduction than GDN in BD-rate sequentially. 
Besides, we confirm the generalization ability on different network depth by modifying the number of cascading layers on ICLR2018\cite{balle2018variational}. In Figure \ref{ablation study}.(c), When adding a layer of cascade, TPM still obtains a 2.24$\%$ reduction in BD-rate compared to GDN. Meanwhile, TPM also achieves a 3.21$\%$ reduction in BD-rate than GDN with one less cascading layer. 
\vspace{-1.5em}

\section{Conclusion}

In this paper, we develop a fundamental new viewpoint to think about the design of transformation in learned image compression by interpreting the learned image compression framework as a communication system.
We first analyzed the consistency between learned image compression and communication system, and then established the correspondence.
Moreover, a transformation method based on modulation techniques  (TSM) is proposed, which can be  generalized to the previous transformation as the linear modulation technique.  By extending to nonlinear  modulation technique, three transformation methods (TPM, TFM, TJM)are obtained.
Furthermore, We design efficient network architectures for above transformations and its residual block   (ResTSM).
Experiment shows that our simplest TPM method outperformed than GDN on four learned image compression  architectures under the same complexity, and our ResTSM method achieves a comparable performance with the SOTA transformation method while the complexity is 29.59\%  of SOTA. 
\appendix
%%%%%%%%% REFERENCES
{\small

\bibliographystyle{ieee_fullname}
%\bibliography{ReviewTemplate}
}

\end{document}